\documentclass[prl,twocolumn,superscriptaddress,showpacs,floats,floatfix]{revtex4}
\usepackage{epsfig,dcolumn,amsmath,latexsym}
\begin{document}

\title{Observation of a parity oscillation in the conductance of atomic wires}

\preprint{7}

\author{R.\,H.\,M.\ Smit}
\affiliation{Kamerlingh Onnes Laboratorium, Leiden University, PO
Box 9504, NL-2300 RA Leiden, The Netherlands}

\author{C.\ Untiedt}
\thanks{Present address: Departamento de F\'\i sica Aplicada,
Universidad de Alicante, Spain. Electronic address: untiedt@ua.es}
\affiliation{Kamerlingh Onnes Laboratorium, Leiden University, PO
Box 9504, NL-2300 RA Leiden, The Netherlands}

\author{G.\ Rubio-Bollinger}
\affiliation{Laboratorio de Bajas Temperaturas, Departamento de F\'\i sica de la Materia Condensada C-III, \\
Universidad Aut\'onoma de Madrid, E-28049 Madrid, Spain}

\author{R.\,C.\ Segers}
\affiliation{Kamerlingh Onnes Laboratorium, Leiden University, PO
Box 9504, NL-2300 RA Leiden, The Netherlands}

\author{J.\,M.\ van Ruitenbeek}
\affiliation{Kamerlingh Onnes Laboratorium, Leiden University, PO
Box 9504, NL-2300 RA Leiden, The Netherlands}

\date{\today}

\begin{abstract}
Using a scanning tunnel microscope or mechanically controllable
break junctions atomic contacts for Au, Pt, and Ir are pulled to
form chains of atoms. We have recorded traces of conductance
during the pulling process and averaged these for a large number
of contacts. An oscillatory evolution of conductance is observed
during the formation of the monoatomic chain suggesting a
dependence on the numbers of atoms forming the chain being even or
odd. This behavior is not only observed for the monovalent metal
Au, as was predicted, but is also found for the other
chain-forming metals, suggesting it to be a universal feature of
atomic wires.
\end{abstract}


\maketitle

Wires of one atom thick and several atoms long (chains of atoms)
connecting two macroscopic electrodes can be formed when pulling
atomic contacts using a scanning tunneling microscope (STM) or
mechanically controllable break junctions (MCBJ)
\cite{Yanson2NAT98}. Chains of atoms have also been formed using a
transmission electron microscope (TEM) where the edges of two
holes in a thin film meet, just before their coalescence
\cite{Ohnishi,Rodrigues,Takai}. Recently it has been shown that
the formation of such structures depends on the metal and that for
clean environments they can only be formed with Au, Pt and Ir
\cite{SmitPRL01,Bahn}.

\begin{figure}[!b]
\epsfig{width=7cm,figure=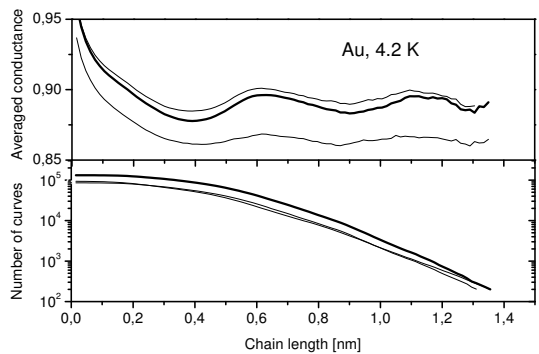} \caption
{\label{fig:Au-curves} Three further examples of measurements of
the averaged conductance plateau for Au at 4.2 K (top panel). The
lower panel shows the number of points used in obtaining the
average and illustrates the rapid decrease in the number of
plateaus for longer lengths, which explains the increase in noise
to the right in the top panel.}
\end{figure}

The atomic chains constitute a unique metallic structure where
one-dimensional properties of matter can be tested. Although they
have been realized only recently, atomic wires have been text book
examples and toy-models for a long time because of their
simplicity. Moreover, they are the ultimate limit in the
miniaturization of electronics, being the simplest object to be
connected into a circuit. This has stimulated numerical
simulations of their transport properties well before their
experimental observation.
Various groups \cite{Lang_Na,Sim,Zeng,Kim,Havu,Gutierrez} have
found oscillations in the conductance as a function of the number
of atoms for calculations of sodium atomic chains, where this
metal was selected because it has the simplest electronic
structure. Sim {\it et al.} \cite{Sim}, using first-principles
calculations and exploiting the Friedel sum rule, found that the
conductance for an odd number of atoms is equal to the quantum
unit of conductance $G_0$ (=$2e^2/h$), independent of the geometry
of the metallic banks, as long as they are symmetric for the left
and right connections. On the other hand, the conductance is
generally smaller than $G_0$ and sensitive to the lead structure
for an even number of atoms. The odd-even behavior follows from a
charge neutrality condition imposed for monovalent-atom wires. The
parity oscillations survive when including electron-electron
interactions \cite{Zeng}, which can even amplify the oscillations
\cite{Molina}. While most authors predict perfect transmission for
chains composed of an odd number of atoms and a transmission
smaller than unity for an even number, some groups find the
opposite \cite{Lang_Na,Havu}. This anomalous phase shift of the
oscillations can likely be attributed the the artificial interface
introduced in going from the Na atoms
to jellium leads as used by those groups \cite{Gutierrez}.\\
\begin{figure}[!b]
\begin{center}
\epsfig{width=6cm,figure=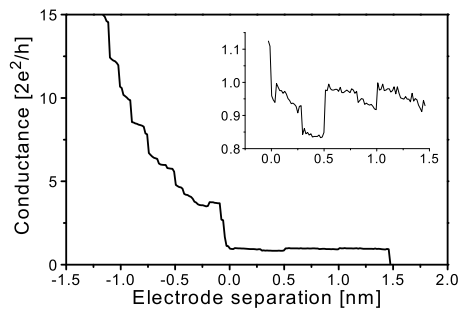} \caption {\label{gold
trace} Evolution of the conductance while pulling a contact
between two gold electrodes. In the inset, an enlargement of the
plateau of conductance at $\sim 1\,G_0$ is shown. Variations to
lower conductance and back up by about 10--15\% can be noticed
when the atomic chain is stretched.}
\end{center}
\end{figure}
Here we present an experimental investigation of the parity effect
by measuring the changes of conductance in the process of pulling
atomic chains of Au, Pt and Ir. We make a statistical analysis in
order to remove the influence of different electrode
configurations and thus uncover the effects which are intrinsic to
the atomic chain. The experiments were performed using MCBJ and
were reproduced for Au with a low-temperature STM. The Au, Pt and
Ir wires have a purity of 99.998\%, 99,998\% and 99.98\%
respectively. In MCBJ the wires are notched and glued on a
substrate that can be bent by use of a piezo element. Once the
sample is at 4.2 K inside a cryogenic vacuum the notch is broken
by bending the substrate resulting in the separation of the wire
into two electrodes with fresh surfaces. The relative displacement
of the two resulting electrodes can be controlled with a
resolution better than a picometer.\\
\begin{figure}[!t]
\epsfig{width=5.5cm,figure=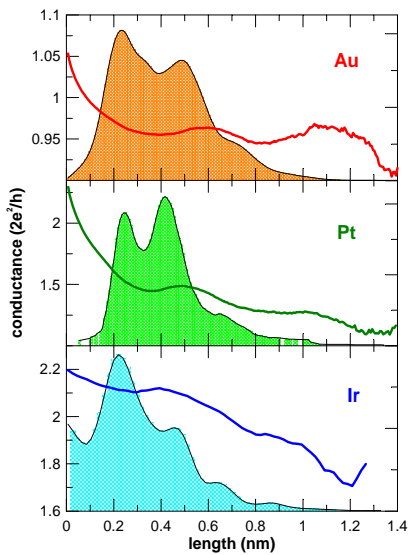} \caption {\label{odd even}
Averaged plateaus of conductance for chains of atoms of the three
metals investigated: Au, Pt, and Ir. Each of the curves are made
by the average of individual traces of conductance while pulling
atomic contacts or chains. Histograms of the plateau lengths for
the three metals obtained from the the same set of data are shown
by the filled curves. They compare well to those in
Ref.\,\protect\cite{SmitPRL01}.}
\end{figure}
When pulling apart two electrodes in contact, the conductance
decreases in a stepwise fashion following the atomic
rearrangements in a succession of elastic and plastic stages
\cite{AgraitPRL95,UntiedtPRB97} (Fig.\,\ref{gold trace}). The
successive values of conductance are related to the sizes of the
contacts between the two electrodes. The last plateau of
conductance before rupture is in general due to a single atom
contact. The formation of an atomic wire results from further
pulling of this one-atom contact, and its length can be estimated
from the length of the last conductance plateau
\cite{Yanson2NAT98,SmitPRL01,UntiedtPRB02}. A histogram made of
those lengths (filled curves in Fig.\,\ref{odd even}) shows peaks
separated by distances equal to the inter-atomic spacing in the
chain. These peaks correspond to the lengths of stretching at
which the atomic chain breaks, since at that point the strain to
incorporate a new atom is higher than the one needed to break the
chain \cite{Gabino}. This implies that a chain of atoms with a
length between the position of the $n^{\rm th}$ and $(n+1)^{\rm
th}$ peak consists typically of $n+1$ atoms.\\
The valence of the metal determines the number of electronic
channels through the chain, and each channel contributes a
conductance with a maximum of the quantum unit, $G_0$ \cite{elke}.
For gold, a monovalent metal, both the one-atom contact and the
chain have a conductance of about $1\,G_0$ with only small
deviations from this value (see Fig.\,\ref{gold trace}) suggesting
that the single channel has a nearly perfect coupling to the
banks. The small changes of conductance during the pulling of the
wire shown in the inset in Fig.\,\ref{gold trace} are suggestive
of an odd-even oscillation. The jumps result from changes in the
connection between the chain and the banks when new atoms are
being pulled into the atomic wire. In order to uncover possible
patterns hiding in these changes we have averaged many plateaus of
conductance starting from the moment that an atomic contact is
formed (defined here as a conductance dropping below 1.2\,$G_0$)
until the wire is broken (conductance dropping below 0.5\,$G_0$).
In the upper panel of Fig.\,\ref{odd even} it can be seen that the
thus obtained average plateau shows an oscillatory dependence of
the conductance with the length of the wire. The amplitude of the
oscillation is small and differs slightly between experiments.
Fig.\,\ref{fig:Au-curves} shows three further examples of similar
measurements for Au out of about a dozen independent experiments.
It shows that the period and phase are quite reproducible.\\
We have repeated the same procedure with the other two metals
forming chains of atoms, namely Pt and Ir. These metals have $s$
and $d$ orbitals giving rise to five channels of conductance. Each
of the channels may have different transmissions that can be
affected by the details of the contact and therefore the average
plateau conductance is expected to show a more complicated
behavior. A one-atom Pt contact has a conductance of about
2\,$G_0$ while for a Pt atomic chain it is slightly smaller,
$\sim1.5\,G_0$ \cite{Soren} with variations during the pulling
process that can be as large as 0.5\,$G_0$. We observe
oscillations similar to those for Au, which we compare to the peak
spacing in the length histogram in Fig.\,\ref{odd even}. The
latter is obtained by taking as a starting point of the chain a
conductance dropping below 2.4\,$G_0$, following
Ref.\,\cite{SmitPRL01}. Ir shows a similar behavior although
somewhat less pronounced and it is more difficult to obtain good
length histograms.\\
The periodicity $p$ of the oscillation in the conductance for the
three metals is about twice the inter-peak distance $d$ of their
corresponding plateau-length histogram as is shown in
table\,\ref{table:periods}.
\begin{table}[!t]
\begin{center}
\begin{tabular}{c|c|c}
Metal & Peak distance $d$ (nm) & Half period $p/2$ (nm)\\
\hline
Au & 0.25$\pm$0.02 & 0.24$\pm$ 0.02\\
Pt & 0.23$\pm$0.02 & 0.25$\pm$ 0.03\\
Ir & 0.22$\pm$0.02 & 0.23$\pm$ 0.04\\
\end{tabular}
\end{center}
\caption {\label{table:periods} Values obtained from the data in
Fig.\,2 for the inter-peak distance $d$ and the oscillation period
of conductance $p$, which is divided by 2 to facilitate
comparison. The period $p$ was obtained from the distance between
the maxima (minima) after subtracting a linear slope. }
\end{table}
This behavior agrees with the expected alternating odd-even
evolution of the conductance with the number of atoms. The phase
of this evolution also agrees with predictions
\cite{Sim,Zeng,Kim,Gutierrez} that chains having an odd number of
atoms should have a higher conductance than even-numbered
chains.\\
This odd-even behavior is essentially an interference effect and
can be easily understood in the frame of a simplified
one-dimensional free-electron model. Consider an atomic wire of a
length $L=n a$ connected to two one-dimensional (1D) reservoirs,
where $n$ is the number of atoms and $a$ the interatomic distance.
The reservoirs are characterized by a Fermi wave vector $k_1$, and
$k_2$ is the equivalent for the 1D channel inside the chain. The
transmission of an electron at the Fermi energy can be obtained
from matching of the wave functions and can be written as
\begin{equation}\label{transmision}
\nonumber
 T=\frac{16
\gamma^2}{(1+\gamma)^4+(1-\gamma)^4-2(1+\gamma)^2(1-\gamma)^2\cos(2
k_2 L)},
\end{equation}
where $\gamma = k_2/k_1$. The connection between wire and
reservoirs is represented by a mismatch in Fermi wave vector and
interference of partial waves scattering from the two contact
points gives rise to the oscillatory term in the denominator. It
is essential to note that $k_2$ is fixed by charge neutrality
inside the wire. An isolated wire of finite length will have a
limited set of accessible states $k_2^i = \frac {\pi}{a}
\frac{i}{n}$ for $i =1,...,n$, that can couple to the reservoirs
when the wire is contacted. The number of electrons will fix the
Fermi energy and the corresponding Fermi wave vector, which in the
case of a mono-valent metal has a value $k_2 = \frac{\pi}{2 a}
(n+1)/n$ After coupling to the reservoirs the Fermi energy may
adjust somewhat, but as chain and reservoir are composed of the
same metal there will be very little charge transfer. Therefore we
obtain $2k_2 L = \pi (n+1)$ that, when substituted in the
expression for $T$ above, gives a transmission $T=1$ for the case
of an odd number of atoms $n$, and $T= 4 \gamma^2 /
(1+\gamma^2)^2$ for $n$ even. We find that for an odd number of
atoms in a chain of a monovalent metal the conductance is $G_0$,
in agreement with Refs.\,\cite{Sim,Zeng,Kim,Gutierrez}. The
conductance for the even-numbered chain is smaller than $1\,G_0$
(since $\gamma<1$) and is sensitive to the interface geometry,
here modelled by $\gamma$.\\
%
Such oscillatory behavior of the conductance is therefore expected
for chains of a monovalent metal such as gold and explains the
observed oscillation and its period and phase. In the averaged
curves of the experiment (Figs.\,\ref{odd even} and
\ref{fig:Au-curves}) the conductance does not quite reach a
maximum of $1\,G_0$. This is largely due to the averaging
procedure, where for a given length there are contributions from
$n$ and $n+1$-numbered chains and only the relative weight of
these varies. In individual traces (Fig.\,\ref{gold trace}) the
maxima come much closer to full transmission. Further suppression
of the maximum conductance may result from asymmetries in the
connections to the leads. The relatively small amplitude of the
oscillations is consistent with the fact that the average
conductance is close to unity, implying that the contact between
the chain and the banks is nearly adiabatic ($\gamma$ is close to
1 in our model). The rise above 1\,$G_0$ for short lengths in
Figs.\,\ref{odd even} and \ref{fig:Au-curves} can be
attributed to tunnelling contributions of additional channels.\\
The fact that similar behavior is also found for Pt and Ir is
unexpected. The above arguments restricting the $k$ vector in the
chains do not apply for multivalent metals and one would expect
contributions from several conductance channels. We have performed
a further test in order to verify whether the oscillations in the
conductance are due to interference in the wave functions. Here we
make use of the curvature of the transmission $T$ as a function of
the applied bias voltage. For odd chains $T$ is at a maximum and
decreases for increasing bias, while the reverse applies for even
chains and the transmission should increase with bias. Therefore,
we have measured simultaneously the conductance and its second
derivative using two parallel lock-in amplifiers tuned to the
first and third harmonic of the applied bias voltage modulation to
the contact, and averaged the signals for a large number of
chains.\\
\begin{figure}[!t]
\epsfig{width=6cm,figure=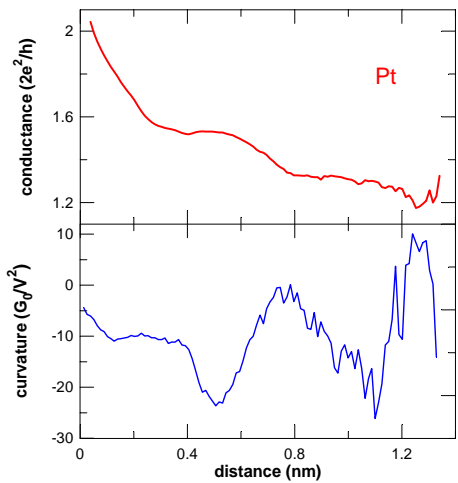} \vspace{0mm} \caption
{\label{curvature} Averaged conductance plateau for chains of Pt
atoms (top panel), measured on a new sample illustrating the
reproducibility of the measurements in Fig.\,\protect\ref{odd
even}. The bottom panel shows the mean value of the second
derivative of the conductance $d^2 G/dV^2$ measured
simultaneously. The data have been obtained using a bias voltage
modulation amplitude of 40\,mV$_{\rm rms}$ at a frequency of
2.37\,kHz and are averaged over 8000 scans.}
\end{figure}
Fig.\,\ref{curvature} shows the results for Pt. The maxima
(minima) in conductance are found to coincide with the minima
(maxima) in the curvature of the conductance $d^2 G/dV^2$, just as
expected from the resonant behavior. Also the increase of the
amplitude if $d^2 G/dV^2$ with the length of the chain agrees with
our simple free electron model, which can be understood from the
fact that the resonances become more closely spaced. This
experiment provides additional evidence for the interpretation of
the oscillations for Pt as an odd-even effect. For Au, because of
the smaller amplitude of the odd-even features, no convincing $d^2
G/dV^2$ signals have been obtained.\\
In addition to the oscillations for Pt and Ir in the mean
conductance of the chains the measurements in Figs.\ref{odd even}
and \ref{curvature} show an unexpected slope of about
0.3--0.4\,$G_0$/nm. For a ballistic wire the conductance as
function of length is expected to be constant, apart from the
oscillatory behavior discussed above. The fact that this decrease
of conductance is found for the multivalent metals and not for Au
suggests that partly-open channels may play a major role. Probably
the overlaps of the electronic states decrease for increasing
numbers of atoms along the chain, which may suggest that only one
or two channels survive for long chains. We would
welcome numerical simulations to test this effect.\\
In summary, we have observed a parity oscillation of the
conductance as a function of the number of atoms forming an atomic
wire. We find this behavior not only for Au, as predicted, but
also for the other two metals that form chains, Pt and Ir,
suggesting that the effect may be universal for atomic wires,
independent of the number of channels. In addition, Pt and Ir show
a monotonic decrease of the conductance with length of the wire on
top of the oscillations which requires further explanation.\\
We thank J.\,C.\ Cuevas, N.\ Agra\"\i t and S.\ Vieira for
stimulating discussions and M.\ Polhkamp for assistance in the
experiments. This work is part of the research program of the
''Stichting FOM''. G.\,R.\ aknowledges suport of the Spanish DGI
by the project MAT 2001-1281. C.U. has been supported by the
European Union under contract No.\ HPMF-CT-2000-00724.\\


\begin{thebibliography}{99}

\bibitem{Yanson2NAT98}A.\,I.\ Yanson, G.\ Rubio-Bollinger, H.\,E.\ van den Brom,
N.\ Agra\"{\i}t, and J.\,M.\ van Ruitenbeek, Nature {\bf 395}, 783
(1998).

\bibitem{Review} N.\ Agra\"{\i}t, A.\ Levy-Yeyati, and J.\,M.\ van
Ruitenbeek, Phys.\ Rep.\ {\bf 377}, 81 (2003).

\bibitem{Ohnishi} H.\ Ohnishi, Y.\ Kondo, and K.\ Takayanagi, Nature {\bf 395},
780 (1998).

\bibitem{Rodrigues} \label{Rod} V.\ Rodrigues, T.\ Fuhrer, and D.\ Ugarte,
Phys.\ Rev.\ Lett.\ {\bf 85}, 4124 (2000).

\bibitem{Takai} Y.\ Takai, T.\ Kawasaki, Y.\ Kimura, T.\ Ikuta, and R.\ Shimizu,
Phys.\ Rev\ Lett.\ {\bf 87}, 106105 (2001).

\bibitem{SmitPRL01} R.\,H.\,M.\ Smit, C.\ Untiedt, A.\,I.\ Yanson, and J.\,M.\ van Ruitenbeek,
Phys.\ Rev.\ Lett.\ {\bf 87}, 266102 (2001).

\bibitem{Bahn} S.\,R.\ Bahn, N.\ Lopez, J.\,K.\ N{\o}rskov and
K.\,W.\ Jacobsen, Phys.\ Rev.\ B {\bf 66}, 081405 (2002).

\bibitem{Lang_Na} N.\,D.\ Lang, Phys.\ Rev.\ Lett. {\bf 79}, 1357 (1997).

\bibitem{Sim} H.\,S.\ Sim, H.\,W.\ Lee, and K.\,J.\ Chang, Phys.\ Rev.\ Lett.\ {\bf 87}, 096803
(2001); H.\,S.\ Sim, H.\,W.\ Lee, and K.\,J.\ Chang, Physica E,
{\bf 14}, 347 (2002).

\bibitem{Zeng} Z.\,Y.\ Zeng and F.\ Claro, cond-mat/0110057;
Z.\,Y.\ Zeng and F.\ Claro, Phys.\ Rev.\ B {\bf 65}, 193405
(2002).

\bibitem{Kim} T.\,S.\ Kim and S.\ Hershfield, Phys.\ Rev.\ B {\bf 65}, 214526 (2002).

\bibitem{Havu} P.\ Havu, T.\ Torsti, M.\,J.\ Puska, and R.\,M.\ Nieminen, Phys.\ Rev.\ B {\bf 66}, 075401
(2002).

\bibitem{Gutierrez} R.\ Guti{\'e}rrez and F.\ Grossmann and R.\ Schmidt, Acta Phys.\ Pol.\ B {\bf 32} 443 (2001).

\bibitem{Molina} R.\,A.\ Molina, D.\ Weinmann, R.\,A.\ Jalabert,
G.-L.\ Igold and J.-L.\ Pichard, Phys.\ Rev.\ B (2003) in print,
cond-mat/0209552.

\bibitem{AgraitPRL95} N.\ Agra\"{\i}t, G.\ Rubio, and S.\ Vieira,
        Phys.\ Rev.\ Lett.\ {\bf 74}, 3995 (1995).

\bibitem{UntiedtPRB97} C.\ Untiedt, G.\ Rubio, S.\ Vieira, and N.\ Agra\"{\i}t,
        Phys.\ Rev.\ B {\bf 56}, 2154 (1997).

\bibitem{UntiedtPRB02} C.\ Untiedt, A.\,I.\ Yanson, R.\ Grande, G.\ Rubio-Bollinger,
N.\ Agra\"{\i}t, S.\ Vieira, and J.\,M.\ van Ruitenbeek, Phys.\
Rev.\ B {\bf 66}, 085418 (2001).

\bibitem{Gabino} G.\ Rubio-Bollinger, S.\,R.\ Bahn, N.\ Agra\"{\i}t, K.\,W.\ Jacobsen, and S.\ Vieira,
Phys.\ Rev.\ Lett.\ {\bf 87}, 026101 (2001).

\bibitem{elke} E.\ Scheer {\it et al.}, Nature {\bf 394}, 154 (1998);
E.\ Scheer, P.\ Joyez, D.\ Esteve, C.\ Urbina, and M.\,H.\
Devoret, Phys.\ Rev.\ Lett. {\bf 78}, 3535 (1997).

\bibitem{Soren}  S.\,K.\ Nielsen {\it et al.} Phys.\ Rev.\ B {\bf 67} 2425XX (2003), in print.

\end{thebibliography}
\end{document}